\documentclass[aps,prl,reprint,groupedaddress]{revtex4-2}
\usepackage{amsmath}
\usepackage{setspace} 
\usepackage{textcomp}
\usepackage{amsfonts}
\usepackage{amssymb}
\usepackage{graphicx}
\usepackage{siunitx}
\usepackage{caption}
\usepackage{setspace}
\usepackage{xcolor}
\usepackage{graphicx}
\usepackage{lipsum}
\usepackage{afterpage}
\usepackage[resetlabels,labeled]{multibib}
\newcites{BC}{BC Readings}
\setcitestyle{super}
\usepackage[colorlinks, citecolor={blue}, linkcolor={black}]{hyperref}

\begin{document}

\title{Control of chirality and directionality of nonlinear metasurface light source via twisting}

\author{Huanyu Zhou$^1$}
\author{Xueqi Ni$^2$}
\author{Beicheng Lou$^3$}
\author{Shanhui Fan$^3$}
\author{Yuan Cao$^1$}
\author{Haoning Tang$^{1,4}$}

\affiliation{$^1$ Department of Electrical Engineering and Computer Science, University of California at Berkeley, Berkeley, CA 94720, USA}
\affiliation{$^2$ Department of Physics, National University of Singapore, Singapore 119077}
\affiliation{$^3$ Department of Applied Physics and Ginzton Laboratory, Stanford University, Stanford, CA 94305, USA}
\affiliation{$^4$ School of Engineering and Applied Sciences, Harvard University, Cambridge, MA 02138, USA}

\begin{abstract}
Metasurfaces have revolutionized nonlinear and quantum light manipulation in the past decade, enabling the design of materials that can tune polarization, frequency, and direction of light simultaneously. However, tuning of metasurfaces is traditionally achieved by changing their microscopic structure, which does not allow \emph{in situ} tuning and dynamic optimization of the metasurfaces. In this Letter, we explore the concept of twisted bilayer and tetralayer nonlinear metasurfaces, which offer rich tunability in its effective nonlinear susceptibilities. Using gold-based metasurfaces, we demonstrate that a number of different singularities of nonlinear susceptibilities can exist in the parameter space of twist angle and interlayer gap between different twisted layers. At the singularities, reflected/transmitted light from the nonlinear process (such as second-harmonic generation) can either become circularly polarized (for C points), or entirely vanish (for V points). By further breaking symmetries of the system, we can independently tune all aspects of the reflected and transmitted nonlinear emission, achieving unidirectional emission with full-Poincar\'e polarization tunability, a dark state (V-V point), or any other bidirectional output. Our work paves the way for multidimensional control of polarization and directionality in nonlinear light sources, opening new avenues in ultrafast optics, optical communication, sensing, and quantum optics.
\end{abstract}

\maketitle

The field of nanophotonics has seen an increasing demand for integrating multiple optical functionalities into a single compact design, for both classical and quantum applications. A crucial component of such a integrated system is a compact and coherent light source \cite{zhang2022spatially,li2018addressable,kruk2015enhanced,kwiat1995new,koshelev2020subwavelength,anthur2020continuous,guo2023ultrathin,santiago2021photon,solntsev2021metasurfaces,marino2019spontaneous,feng2024polarization}.Because coherent sources are readily available only in certain frequency bands, frequency conversion via nonlinear optical effects is an important ingredient in integrated light sources. Progress in this field relies on the development of novel nonlinear materials \cite{yao2021enhanced,kim2024three,sutherland2003handbook,boyd2008nonlinear,wu2017giant,li2013probing,guyot1987local,tang_-chip_2024,ha2021enhanced,trovatello2024tunable} and nonlinear metasurfaces \cite{li_nonlinear_2017,krasnok2018nonlinear,kolkowski2023nonlinear,sharma2023electrically,segal2015controlling,litchinitser2015optical,yu2022electrically,ren2020tailorable,celebrano2015mode,martorell1997second,minkov2019doubly,lapine2014colloquium,lee2014giant,linnenbank2016second,du2020twisting,mann2023inverse}. 
The layers of these materials can be further engineered to enhance their optical nonlinearity, e.g. by quasi-phase-matching \cite{myers1995quasi}.

However, the practical application of these materials is hindered by the lack of wide tunability towards handedness and emission directions. To the best of our knowledge, there is no efficient on-chip strategy to simultaneously control both the circular dichroism of the nonlinearity (difference between the Left-handed Circular Polarization (LCP) and Right-handed Circular Polarization (RCP) components of the generated nonlinear light, which in turn controls its polarization) and the directional dichroism of the nonlinearity (difference between nonlinear light that propagate upwards or downwards) to date \cite{ni2024three,yin_observation_2020,yin2023topological,gong2023multipolar,krasnok2018nonlinear,raval2017unidirectional}. One major reason is that the state-of-the-art nonlinear metasurfaces lack \emph{in situ} tunable degrees of freedom (DoFs), as their tunabilities typically requires changing the underlying meta-atoms.

Twisted bilayer (or multilayer) metasurfaces possess additional interlayer DoFs, such as twist angle ($\theta$) and interlayer gap ($h$) \cite{tang2021modeling,tang2022chip,tang2023experimental,lou2022tunable,lou2021theory,raun2023gan,tang2023chip,nguyen2022magic,li2024twisted,wang2023twist,qin2023arbitrarily,zheng2022molding,hu2020topological,yves2024twist,hu2021twistronics,askarpour2014wave,hu2020moire,lou2024free,huang2022moire,yi2022strong,dong2021flat,chen2021perspective,mao2021magic,oudich2021photonic,hong_twist_2023,zhang_twisted_2023}. By carefully designing the meta-atom structures, optical properties such as frequency, polarization, beam angle, and phase can be widely and simultaneously tuned through nanometer-scale displacement between layers of near-field coupled metasurfaces. The twisted bilayer (or multilayer) nonlinear metasurfaces have their nonlinear susceptibilities determined by the macroscopic susceptibility of the meta-atoms, in addition to the interlayer DoFs. Moreover, MEMS-nanophotonic technology \cite{tang_-chip_2024,tang2023chip} enables dynamic adjustment of interlayer DoFs, allowing for fully in-situ tuning of the nonlinear susceptibility.

In this Letter, we study the mechanisms for tuning the chirality and directionality of the second-order effective susceptibility ($\chi^{(2)}$) in the twisted bilayer and tetralayer nonlinear metasurfaces. We first consider a simple case with two layers with parameters $\theta$ (twist angle) and $h$ (interlayer spacing). In this space, we find topological singularities C points for chiral $\chi$ \cite{zhen2014topological,hsu2016bound,chen2019singularities,liu_circularly_2019,yoda_generation_2020,yin_observation_2020,yin2023topological,zeng_dynamics_2021,ye2020singular,guo2020meron,doeleman2018experimental,chen2020line,otte2018polarization}. By slightly tuning the meta-atom structure, we can achieve also a V point (vortex $\chi$) singularity where the nonlinear susceptibility for both chiralities vanishes. Furthermore, we demonstrate that independent control of all four components of susceptibilities (two chiralities multiplies two directions) can be achieved in twisted tetralayer metasurfaces by tuning the various $h$ and $\theta$ between layers, without a need to change the meta-atom structure in this case. Our results provide new insights into tailoring nonlinear optical emission in nanoscale devices, with applications such as chiral laser sources \cite{hickstein2015non}, telecommunication multiplexing \cite{ma2019optical}, quantum information processing \cite{zhang2022all}, sensing and imaging \cite{schlickriede2020nonlinear,kruk2022asymmetric}, optical switching and modulation \cite{klimmer_all-optical_2021,yu2022electrically,wang_all-optical_2024}, entangled light generation \cite{marino2019spontaneous,zhang2022spatially,santiago2022resonant}, and nonlinear holography \cite{ye2016spin,reineke2019silicon}. 

\begin{figure}[!ht]
\centering
\includegraphics[width=\linewidth]{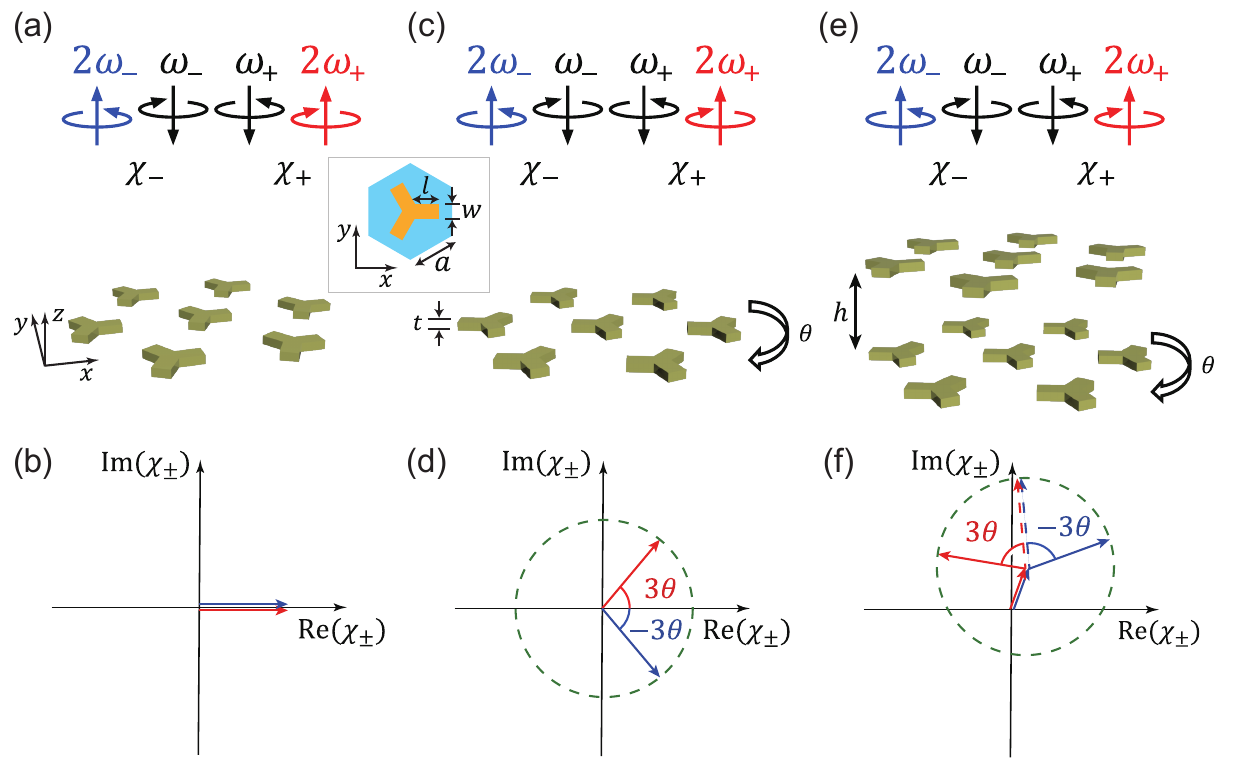}
\caption{\textbf{Effective second order nonlinear susceptibility ($\chi^{(2)}$) in single-layer and twisted bilayer metasurfaces.} (a) The SHG process of a single-layer gold metasurface with $C_{3v}$ symmetry. Black arrows: input light with LCP and RCP component. Red (Blue) arrow: upward output SHG with LCP (RCP) component. Subfigure: Geometric parameters of the metasurface. (c) The SHG process in twisted single-layer metasurface, where the twist angle relative to the $x$-axis is $\theta$. (e) The SHG process in twisted bilayer metasurfaces. The two layers are separated by an interlayer gap $h$, and the second layer has a twist angle of $\theta$ against the first layer. (b,d,f) The $\chi_{\mathrm{eff}}$ phasors that represent the contributions from the LCP (RCP) chiral basis of untwisted single-layer (b), twisted single-layer (d) and twisted bilayer (f) metasurface(s). The dashed circle shows the range that $\chi_+$ and $\chi_-$ can reach when $\theta$ varies.} 
\label{fig:fig1}
\end{figure}

We first consider a single-layer metasurface, as shown in Fig. \ref{fig:fig1}a. The metasurface consists of a lattice of gold meta-atoms with three-fold symmetry ($C_{3v}$), with one of the arms of the meta-atom aligned with the $x$-axis. Each meta-atom is defined by periodicity $a$, thickness $t$, arm length $l$, and arm width $w$. (Fig.\ref{fig:fig1} Subfigure)
Due to the $C_{3v}$ symmetry, its second-order susceptibility $\chi^{(2)}$ has only one nonzero component,
$\chi_{0}\equiv\chi_{xxx}=-\chi_{yxy}=\chi_{yyx}=\chi_{xyy}$ \cite{boyd2008nonlinear}. 
Considering an SHG process, the two components of $\mathbf{E}^{2\omega}$ of the SHG electric field can be written as, 
\begin{equation}
\label{eq:eq1}\begin{bmatrix} E^{2\omega}_{x} \\ E^{2\omega}_{y} \end{bmatrix}
\propto \chi_{0}\begin{bmatrix} (E^{\omega}_{x})^2-(E^{\omega}_{y})^2 \\ -2E^{\omega}_{x}E^{\omega}_{y} \end{bmatrix},
\end{equation}
where $E_{x,y}^{\omega}$ are the components of the fundamental electric field $\mathbf{E}^{\omega}$.

It is more insightful if we decompose $\mathbf{E}^{2\omega}$ into the chiral bases as $\mathbf{E}^{2\omega} = \frac{1}{\sqrt{2}} E^{2\omega}_{+}(\hat{\mathbf{x}} + i\hat{\mathbf{y}}) + \frac{1}{\sqrt{2}} E^{2\omega}_{-}(\hat{\mathbf{x}} - i\hat{\mathbf{y}})$, and similarly for $\mathbf{E}^{\omega}$. Then Eq. \ref{eq:eq1} becomes $E^{2\omega}_{\pm} \propto \sqrt{2}\chi_{0}(E^{\omega}_{\mp})^2$, which implies a well-known conclusion for $C_{3v}$-symmetric nonlinear materials: they have fully decoupled second-order nonlinear response for LCP and RCP light: an LCP fundamental wave always generate an RCP SHG response, and \emph{vice versa} \cite{li_nonlinear_2017} (see Fig. \ref{fig:fig1}a). 
Note that the reflected SHG light acquires an extra flip in chirality as the propagation direction is flipped. Therefore, the SHG electric field in the upward (reflected) direction can be rewritten as (See Supplementary Information Section 1 for the downward (transmitted) wave):
\begin{equation}
\label{eqn:lb1}
    E^{2\omega}_{\pm} \propto \chi_{\pm}(E^{\omega}_{\pm})^2
\end{equation}
where $\chi_{\pm}=\sqrt{2}\chi_{0}$ is the effective nonlinear susceptibility written in the chiral basis. In general, $\chi_{\pm}$ are complex numbers that can be represented as phasors on the complex plane (see Fig. \ref{fig:fig1}b-f).
Rotating the single-layer metasurface by $\theta$ (Fig. \ref{fig:fig1}c) will alter the $\chi_{ijk}$ tensor according to tensor transformations (see Supplementary Information Section 1),
and the effective nonlinear susceptibility $\chi_\pm$ is modified as
\begin{equation}
\chi_{\pm} = \sqrt{2}\chi_{0} e^{\pm3i\theta}.
\end{equation}
As a result, the corresponding phasors is rotated by $\pm3\theta$ on the complex plane (See Fig. \ref{fig:fig1}d).

Before delving into multilayer metasurfaces, we briefly comment on an intuitive understanding of $\chi_{\pm}$ and its visualization throughout this Letter. The spinor $[\chi_+, \chi_-]^\mathrm{T}$, when normalized, can be thought as denoting a point on the Poincar\'e sphere. One could show that this point exactly describes the polarization state (generally elliptical) of the SHG wave, when the fundamental wave is horizontally (\emph{i.e.} $x$-) polarized (see Supplementary Information Section 3). Therefore, without loss of generality, we could visualize $\chi_{\pm}$ at anywhere in the parameter space as a polarization ellipse, which would be the measured SHG polarization when excited by a $x$-polarized fundamental wave.

Now we stack one layer of metasurface on top of the another with a separation of $h$ and a twist of $\theta$ (See Fig. \ref{fig:fig1}e). Since the perpendicular symmetry is broken, we now differentiate between upward nonlinear susceptibility for reflected SHG wave, $\chi_{\mathrm{up},\pm}$, and the downward nonlinear susceptibility for transmitted SHG wave, $\chi_{\mathrm{down},\pm}$. 

As the top layer is untwisted and the bottom layer is twisted by $\theta$, the upward and downward susceptibilities satisfy
\begin{align}
\chi_{\mathrm{up},\pm} &=\chi_{\mathrm{up},1, \pm}+\chi_{\mathrm{up},2, \pm}e^{\pm3i\theta}, \\
\chi_{\mathrm{down},\pm} &=\chi_{\mathrm{down},1, \pm}+\chi_{\mathrm{down},2, \pm}e^{\mp3i\theta},
\end{align}
respectively, where $\chi_{\mathrm{up/down}, j, \pm}$ are the upward/downward susceptibilities of $j$-th layer ($j=1,2$). In the phasor picture (Fig. \ref{fig:fig1}f), these correspond to vector sum of phasors from each layer, with the phasors of the bottom layer rotated by $\pm3\theta$ on the complex plane. It should be noted that $\chi_{\mathrm{up/down},j,\pm}$ are complex functions of $h$ and the geometry of the meta-atoms and are determined by a combination of transfer matrix method and COMSOL simulation (see Supplementary Information Section 2).

\begin{figure}[!ht]
\centering
\includegraphics[width=\linewidth]{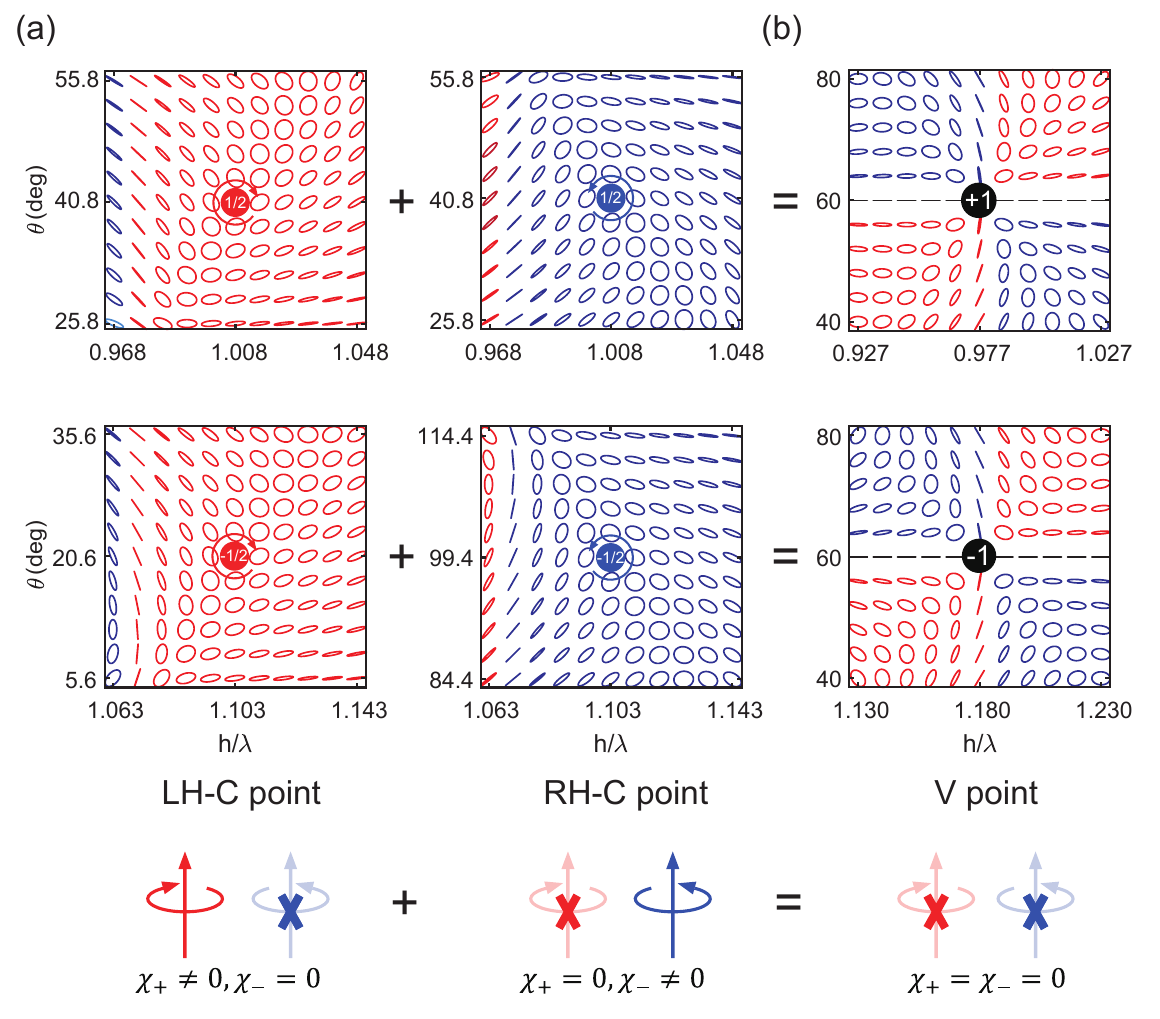}
\caption{\textbf{Parameter space singularities of $\chi_{\mathrm{up},\pm}$ in twisted bilayer metasurfaces.} (a) Four types of C points with different winding textures that are described by the vorticity $v$ and the handedness (left-handed (LH) and right-handed (RH)). LH (RH)-C points are denoted as red (blue) dots at the center of the subfigures.The vorticities are labeled on the dots. (b) A singularity with integer vorticity (V point) can be achieved by merging two C points with identical half-integer vorticity and opposite handedness.} 
\label{fig:fig2}
\end{figure}

The effective nonlinear susceptibility in twisted bilayer metasurfaces has topological features that can be highly designed. 
In Fig. \ref{fig:fig2}a-b, we show the effects of tuning $h$ and $\theta$ on $\chi_{\mathrm{up},\pm}$ (visualized as polarization ellipses, see comments above), exhibiting different types of topological singularities.
Here we used gold meta-atoms with $a=370\mathrm{nm}$, $t=30\mathrm{nm}$, and $w=50\mathrm{nm}$. Here we also vary the arm length $l$ to obtain additional tunability on the susceptibility.
The gold meta-atoms are embedded in glass with an index $n=1.5$, excited with a fundamental wavelength of $\lambda_0=1220\mathrm{nm}$ in vacuum. The actual wavelength in glass is defined as $\lambda=\lambda_0/n$.
$\chi_{\mathrm{up},\pm}$ is periodic in both $h$ and $\theta$, with periods of $\lambda/2$ (due to the Fabry-Pérot oscillation) and $120^\circ$ (due to the lattice $C_{3v}$ symmetry), respectively. This means the $h-\theta$ synthetic space is topologically equivalent to a 2D Brillouin zone. 

As shown in Fig. \ref{fig:fig2}a-b, rich features exist at certain locations of the $h$-$\theta$ parameter space, including "C" points (Chiral $\chi$) and "V" points (Vortex $\chi$). These topological textures are similar to the polarizational C points (circular polarized points) and V points (vortex center points) in photonic crystals \cite{zhen2014topological,hsu2016bound,chen2019singularities,liu_circularly_2019,yoda_generation_2020,yin_observation_2020,yin2023topological,zeng_dynamics_2021,ye2020singular,guo2020meron,doeleman2018experimental,chen2020line,otte2018polarization}. These singularities can be described by their vorticity \cite{zhen2014topological,yoda_generation_2020,zeng_dynamics_2021}:
\begin{equation}
    v=\frac{1}{2\pi}\oint_L \mathrm{d}\mathbf{r} \cdot \nabla_\mathbf{r} \phi(\mathbf{r})
\end{equation}
where $\mathbf{r}=(h,\theta)$ is position vector in parameter space, $L$ is a closed loop encircles the singularity counterclockwise in parameter space, and $\phi(\mathbf{r})$ is the azimuthal angle of the major axis of the polarization ellipse. 

A C point is characterized by a half-integer vorticity ($v = \pm 1/2$). At the C point, the azimuthal angle of major axis of the polarization ellipse becomes ill-defined because it is perfectly circularly polarized. Apart from the vorticity, a C point also has two possible handedness, one with $\chi_{\mathrm{up},-}=0$ (LH-C point) and one with $\chi_{\mathrm{up},+}=0$ (RH-C point).There are thus four types of C points in total from the combination of vorticity and handedness. C points always appear in pairs because each carries a half-integer topological charge \cite{liu_circularly_2019,yoda_generation_2020,chen2020line,yin_observation_2020,ye2020singular,guo2020meron,zeng_dynamics_2021}. As an example, when $l=86.0\mathrm{nm}$ for the top layer and $l=84.0\mathrm{nm}$ for the bottom layer, two LH-C points with $v=\pm 1/2$ (red) and two RH-C points with $v=\pm 1/2$ are found in the parameter space, shown in Fig. \ref{fig:fig2}a. 
 At LH (RH)-C points, the Poincar\'e vector of $[\chi_{\mathrm{up},+},\chi_{\mathrm{up},-}]^\mathrm{T}$ points towards $+z$ ($-z$) directions, and since $\chi_-$ or $\chi_+$ becomes zero, LCP (RCP) SHG is always produced regardless of the input polarization state.

Interestingly, by varying the arm length $l$ of the top layer, we could coalesce a pair of C points into a single V point. Two C points with opposite handedness and identical vorticity can merge, into a V point with integer vorticity ($v=\pm 1$)  \cite{liu_circularly_2019,yoda_generation_2020,chen2020line,yin_observation_2020,ye2020singular,guo2020meron,zeng_dynamics_2021} (see Fig. \ref{fig:fig2}b).
At a V point, nonlinear susceptibility of both chiralities vanish, \emph{i.e.} $\chi_{\mathrm{up},+}=\chi_{\mathrm{up},-}=0$, and the twisted metasurfaces can \emph{no longer emit any SHG wave upward}. In the example shown in Fig. 2, when we tune the arm length $l$ of the top layer to be $83.4\mathrm{nm}$ (up) or $79.7\mathrm{nm}$ (down), different pairs of C points coalesce into a V points in each situation. For twisted bilayer metasurfaces, V points only appear on high-symmetry twist angles, \emph{i.e} when $\theta$ is a multiple of $60^\circ$.

\begin{figure}[!ht]
\centering
\includegraphics[width=\linewidth]{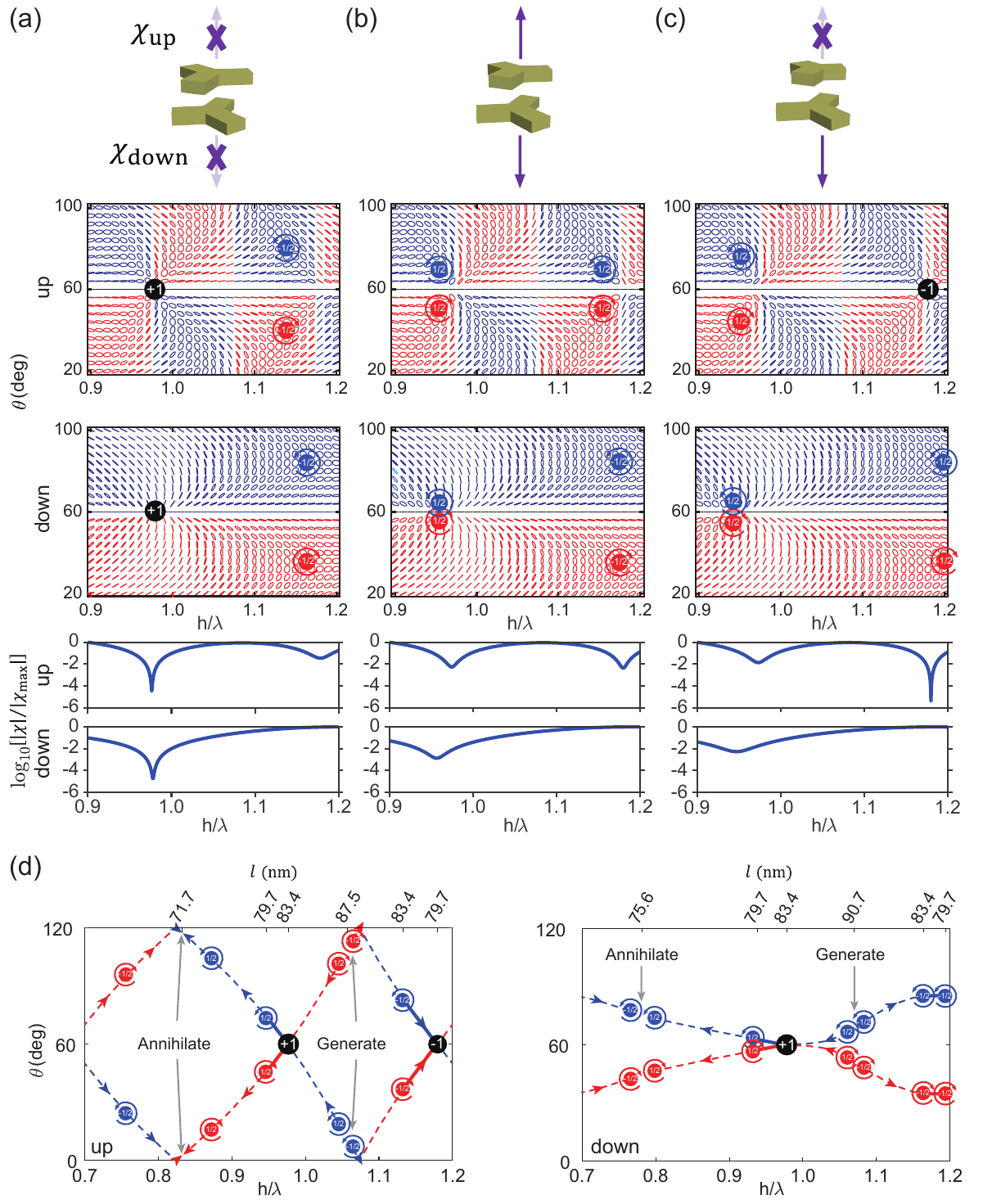}
\caption{\textbf{Evolution of singularities in twisted bilayer metasurfaces.} (a-c) Evolution of singularities in parameter space and relative amplitude of $\chi$ on $\theta=60^\circ$ line when (a) $l=83.4\mathrm{nm}$, (b) $l=81.0\mathrm{nm}$ and (c) $l=79.7\mathrm{nm}$. (a) V-V point at $h = 0.977 \lambda$ and $\theta = 60^\circ$, where V points with vorticity $+1$ exist at both upward and downward directions. (b) Integer singularities on both side split to two half-integer singularities. (c) V-L point at $h=1.198 \lambda$ and $\theta = 60^\circ$, where another pair of half-integer singularities with $v= -1/2$ merge into a new V point with $v=-1$ for the upward direction but remain separated for downward direction. (d) Trajectories of the half-integer singularities in the parameter space as the arm length of upper layer varies. The range of upper layer $l$ in (a)-(c) is shown in solid lines. For the upward (downward) direction, the pair of C points are annihilated at $l=71.7\mathrm{nm}$ ($l=75.6\mathrm{nm}$) and are generated at $l=87.5\mathrm{nm}$ ($l=90.7\mathrm{nm}$).}
\label{fig:fig3}
\end{figure}

In Fig. \ref{fig:fig3}, we study the evolution of the C and V points in both upward and downward nonlinear susceptibilities, as $l$ of the top layer is continuously varied. We start with a structure with the arm length of top layer $l=83.4\mathrm{nm}$ and that of the bottom layer $l=84.0\mathrm{nm}$, as shown in Fig. \ref{fig:fig3}a. In this configuration, the parameter space possesses a V point and a pair of C points in both directions. Here the V points of upward and downward directions exist at the same position in the parameter space $(h,\theta)=(0.977\lambda, 60^\circ)$, and we shall call it a V-V point.
At this point, $\chi_{\mathrm{up},+}=\chi_{\mathrm{up},-}=\chi_{\mathrm{down},+}=\chi_{\mathrm{down},-}=0$, which means the SHG amplitudes vanish in both directions regardless of the input polarization (the SHG power in both directions is reduced by more than five orders of magnitude). This can be viewed as a perfect ``phase-mismatching'' condition where SHG waves from the two metasurfaces fully cancel each other.

When the top layer $l$ is decreases to $81.0\mathrm{nm}$, V points of both directions split into pairs of C points. Therefore, the SHG amplitudes in both directions are now non-zero everywhere in the $h-\theta$  space. (Fig. \ref{fig:fig3}b)
We note that such pairs of C points are always mirror-symmetric with respect to the line of $\theta=60^\circ$, which is typically called the L line\cite{liu_circularly_2019,yoda_generation_2020,chen2020line,yin_observation_2020,zeng_dynamics_2021} because the SHG is always linearly polarized on this line when the fundamental light is linear polarized. 

When the top layer $l$ is further decreased to $79.7\mathrm{nm}$, the initially split C points with opposite handedness and vorticity ($v=-1/2$) merge at $(h, \theta) = (1.180\lambda, 60^\circ)$, and this happens only for $\chi_{\mathrm{up}}$ but not for $\chi_{\mathrm{down}}$. (see Fig. \ref{fig:fig3}c) In other words, at this L-line point, $\chi_{\mathrm{up},+}=\chi_{\mathrm{up},-}=0$, but $\chi_{\mathrm{down},+}=\chi_{\mathrm{down},-}\neq0$, which we call a V-L point. We have thereby achieved unidirectional SHG emission: the SHG amplitude in the upward direction is zero while SHG amplitude in the downward direction is non-zero (their amplitude at least differs by five orders in our calculations).

A full diagram of the trajectories of the singularities is shown in Fig. \ref{fig:fig3}d. The vorticity is conserved in the generation, annihilation, and coalescence of the C points.

\begin{figure}[!ht]
\centering
\includegraphics[width=\linewidth]{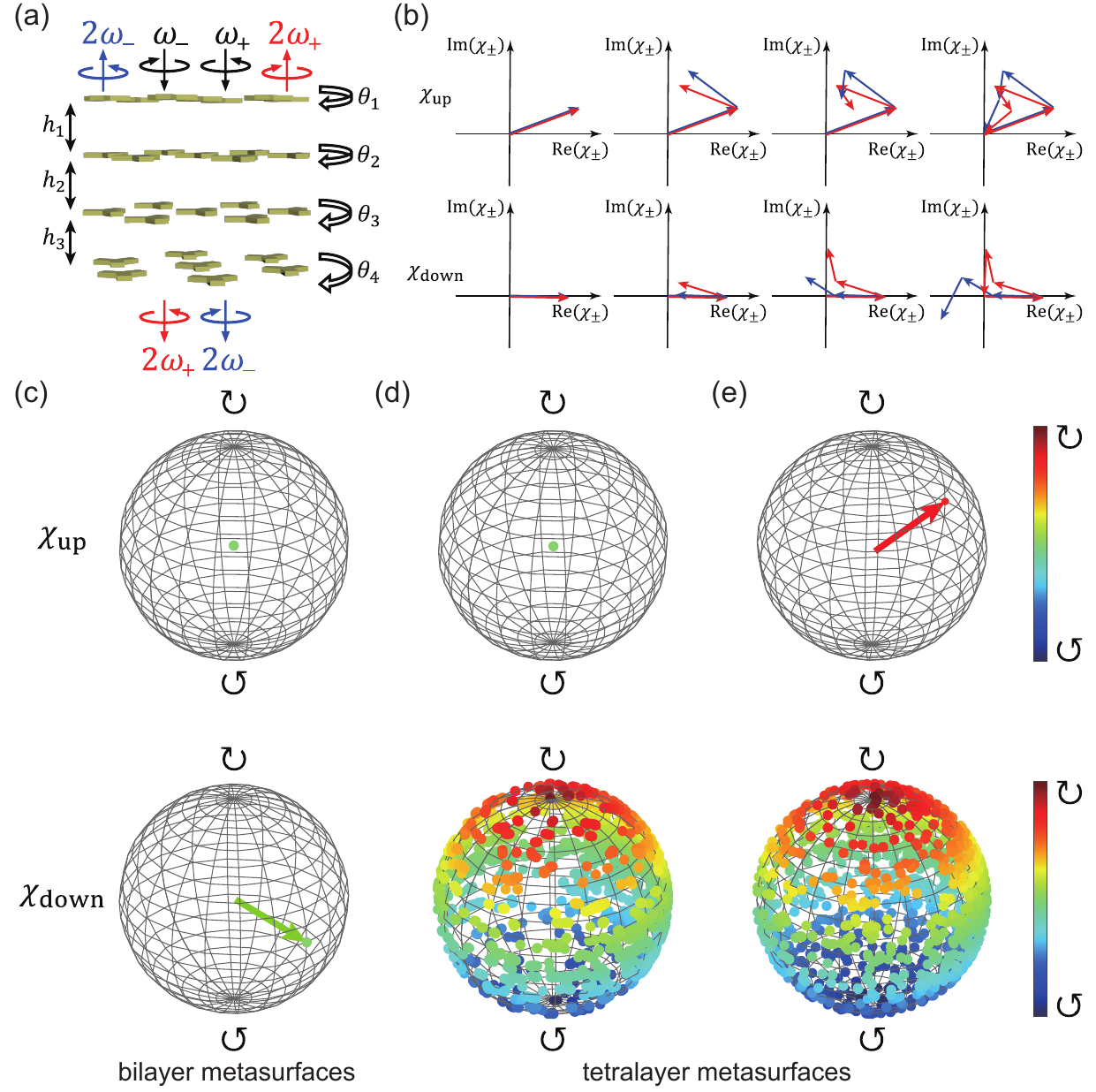}
\caption{\textbf{Singularities of $\chi$ in twisted tetralayer metasurfaces.} (a) The SHG process of the twisted tetralayer metasurfaces. (b) Phasors of $\chi_{\mathrm{up/down}}$ for twisted tetralayer metasurfaces with V point in upward direction and RH-C point in downward direction. (c) In the bilayer metasurfaces, when $\left[\chi_{\mathrm{up},+}, \chi_{\mathrm{up},-}\right]^\mathrm{T} = \left[0, 0\right]^\mathrm{T}$, $\chi_{\mathrm{down}}$ must satisfies $\chi_{\mathrm{down},+}=\chi_{\mathrm{down},-}$, corresponding to $[1,0,0]$ on the Poincar\'e sphere. (d) In the tetralayer metasurfaces, when $\left[\chi_{\mathrm{up},+}, \chi_{\mathrm{up},-}\right]^\mathrm{T} = \left[0, 0\right]^\mathrm{T}$, $\chi_{\mathrm{down}}$ can be manipulated arbitrarily and covers the whole Poincar\'e sphere by varying $h_j$ and $\theta_j$. (e) Similarly when $\left[\chi_{\mathrm{up},+}, \chi_{\mathrm{up},-}\right]^\mathrm{T} \propto \left[2+i, 1\right]^\mathrm{T}$, $\chi_{\mathrm{down}}$ can also be manipulated arbitrarily and covers the whole Poincar\'e sphere by varying $h_j$ and $\theta_j$.}
\label{fig:fig4}
\end{figure}

We now turn to the case with more than two metasurface layers. Multilayer twisted metasurfaces provide additional degrees-of-freedom for tuning $\chi$, and changing the structure of individual meta-atoms is no longer necessary for achieving the same or more tunability than those described above. 
Figure \ref{fig:fig4}a shows the schematic of a twisted tetralayer metasurfaces with separations of $h_{1,2,3}$ and twist angles of $\theta_{1,2,3,4}$ between each layer. Among them, $\theta_1$ can be fixed to zero without loss of generality.
The effective $\chi$ in upward and downward directions for tetralayer metasurfaces can be similarly formulated as,
\begin{equation}
    \chi_{\mathrm{up},\pm}=\sum_{j=1}^N \chi_{\mathrm{up},j,\pm} e^{\pm3i\theta_j}, \chi_{\mathrm{down},\pm}=\sum_{j=1}^N \chi_{\mathrm{down},j,\pm} e^{\mp3i\theta_j}.
\end{equation}
With more tuning DoFs, the effective $\chi$ in different directions in tetralayer metasurfaces can be more freely designed. 
As shown in Fig. \ref{fig:fig4}b, the phasor for each of the four components (up/down, $\pm$) of the effective $\chi$ is represented by a chain of arrows connected head-to-tail, corresponding to amplitude and phase accumulation given by each layers. If an arrow routes back to the origin, then the corresponding component of $\chi$ is zero, and a singularity appears. In this case, we see three phasors routing back to the origin, indicating that $\chi_{\mathrm{up},+}=\chi_{\mathrm{up},-}=\chi_{\mathrm{down},-}=0$, while $\chi_{\mathrm{down},+}\neq0$. This is thus a V-C point, which means a V point for the upward direction and a C point for the downward direction. 

In fact, four layers of metasurfaces are sufficient to synthesize any combination of upward and downward SHG output except a V-V point. Each of the four components of $\chi$ requires two DoFs to independently tune its real and imaginary part. Since an overall phase on the spinor $[\chi_{\mathrm{up/down},+},\chi_{\mathrm{up/down},-}]$ does not have an observable effect and we consider the amplitude as normolized, two DoFs are required to fully control the polarization of SHG in one direction except a V point which requires four DoFs. As a result, six DoFs are sufficient to fully control SHG emission in both directions including a V point in one direction ($4+2=6$), except V points in both directions which requires totally eight DoFs.
The tetralayer metasurfaces have six maneuverable DoFs (three $h_j$ and three $\theta_j$), and are thus the minimal structure capable of synthesizing arbitrary bidirectional nonlinear output except V-V point, which needs pentalayer metasurfaces to provide eight maneuverable DoFs. 

Compared to the bilayer counterpart, tetralayer structure has several advantages. Firstly, the arm length of meta-atoms no longer needs to be modified. Secondly, the unidirectional SHG emissions are no longer limited to the V-L points, where downward emission is limited to linear polarizations when the fundamental light is linear polarized (upward emission is zero). We further elaborate the latter point in Fig. \ref{fig:fig4}c-e, which shows the the spinor representation of $\chi_{\mathrm{up/down},\pm}$ on the Poincar\'e sphere. Fig. \ref{fig:fig4}c corresponds to the V-L point in the bilayer case (Fig. \ref{fig:fig3}c), where the downward emission is limited to the equator (linear polarizations). On the other hand, Fig. \ref{fig:fig4}d shows that in tetralayer structures, full-Poincar\'e polarization tunability can be realized in the downward emission, while the upward emission is kept at zero. Fig. \ref{fig:fig4}e shows that the same full-Poincar\'e tunability in the downward emission can still be achieved when the upward emission is nonzero and fixed to a pre-defined amplitude and polarization. This last example demonstrates the capability to reach arbitrary upward amplitude/polarization and downward amplitude/polarization using all six available DoFs.

In conclusion, twisted tetralayer metasurfaces offer a high degree of tunability for tailoring the effective $\chi$ through the engineering of optical singularities. Our results are generic and applicable to any second-order nonlinear processes, including spontaneous parametric down-conversion (SPDC) and Sum-frequency generation (SFG), and potentially third-harmonic generation (THG) if the $C_{3v}$ metastructure is replaced with a $C_{4v}$ one\cite{li_nonlinear_2017}. The same concepts are also applicable to other nonlinear materials such as 2D materials\cite{tang_-chip_2024}.

\section*{Acknowledgements}
S.F. and B.L. acknowledges support from the U.S. Air Force Office of Scientific Research (Grant No. FA9550-21-1-0244), and from the U.S. Office of Naval Research (Grant No. N00014-20-1-2450).
The authors declare no competing financial interest.

\bibliographystyle{apsrev4-2}
\bibliography{multilayerSHGref.bib}

\end{document}